\documentclass[%
 reprint,
 amsmath,amssymb,
 aps,
pre,
]{revtex4-2}

\usepackage{comment}
\usepackage{graphicx}% Include figure files
\usepackage{dcolumn}% Align table columns on decimal point
\usepackage{bm}% bold math
\usepackage{appendix}
\usepackage{dsfont}
\usepackage{cancel}

\usepackage{xcolor}

\begin{document}

\preprint{APS/123-QED}

\title{Extreme events at the onset of epileptic-like intermittent activity of FitzHugh-Nagumo oscillators on small-world networks}

\author{Javier Cubillos Cornejo}
 \email{javier.cubillos@ug.uchile.cl}
\author{Miguel Escobar Mendoza}
\author{Ignacio Bordeu}%
 \email{ibordeu@uchile.cl}
\affiliation{%
Departamento de Física, Facultad de Ciencias Físicas y Matemáticas, Universidad de Chile
}%

\begin{abstract}
In this work, we explore the influence of coupling strength, network size, and randomness on the collective dynamics of FitzHugh-Nagumo oscillators on complex networks. Using Watts-Strogatz small-world network connectivities, we identify four distinct dynamical phases: chaotic, intermittent, partially synchronized, and fully synchronized. The intermittent phase is characterized by the coexistence of chaotic behavior and chimera states, reminiscent of epileptic-seizure-related (ESR) intermittency observed in the brain. We analyze the inter-spike intervals of the individual oscillators, and the existence, duration, and frequency of ESR events as a function of the system parameters. Furthermore, we study the transitions into and out of the intermittent phase and show that peaks in the probability of extreme events—short transients of anomalously high synchronization—precede the transitions from chaos to intermittency and from partial to full synchronization. These transitions are followed by significant changes in the maximum Lyapunov exponent and Kaplan-Yorke dimension. 
Finally, we discuss how the coupling strength and network properties can be leveraged to control the system's state and the potential applications of extreme event analysis in the study of neural data.
\end{abstract}

\maketitle

\section{\label{sec:Intro}Introduction}
\begin{figure}[t!]
\includegraphics[width=0.5\textwidth]{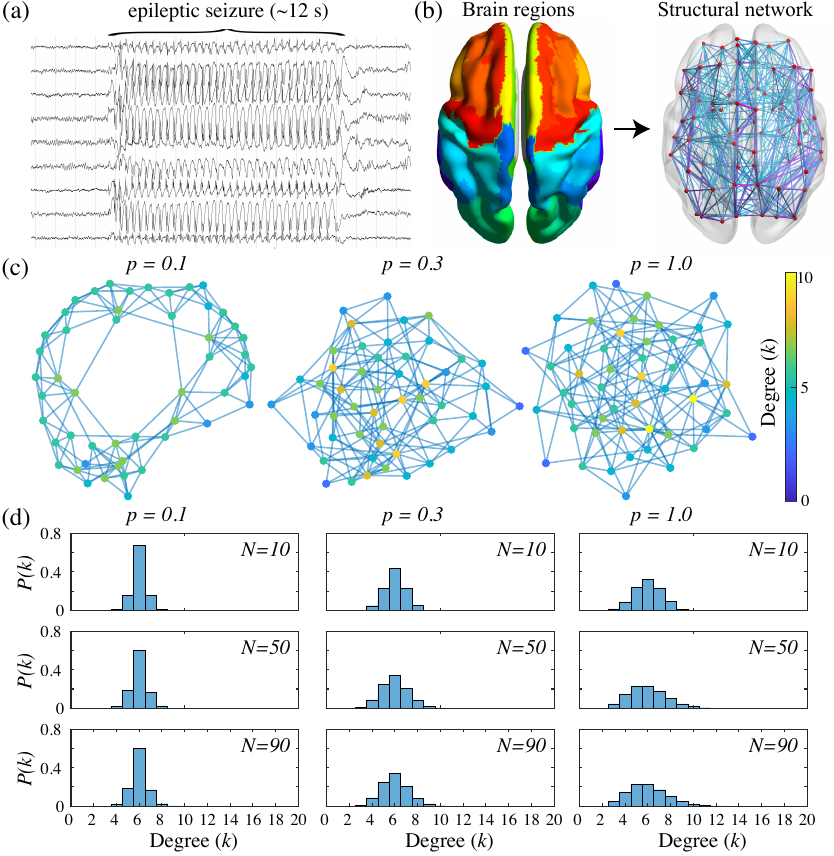}
\caption{\label{fig:motivation} \textbf{Network representation of the brain}. (a) Electroencephalogram (EEG) from a human brain exhibits synchronized regular activity between distinct regions of the brain, characteristic of an epileptic seizure event (adapted from \cite{cerminara2012two}). (b) Anatomical regions of the brain (left). Regions are interpreted as nodes, and the connectivity network (right) is inferred from the neural activity (adapted from \cite{sun2015progressive}). (c) WS networks for different values of the rewiring probability ($p$), color of the nodes indicate their degree ($k$).
(d) Degree distribution for WS networks for different values of $p$ (increasing from left to right), and different network sizes (increasing from top to bottom).} 
\end{figure}

Synchronization is a ubiquitous natural phenomenon, observed across diverse fields, ranging from physics and chemistry to biology, neuroscience, and even social systems \cite{pikovsky2001universal, kuramoto1984chemical, boccaletti2018synchronization, pruessner2012self, barabasi2013network}. In the brain, coordinated firing of distinct regions of the brain is crucial for effective information processing. However, when dysregulated, synchronized states can trigger epileptic seizures (see Fig.~\ref{fig:motivation}a), which disrupt normal brain function \cite{sadleir2008factors, bonifazi2009gabaergic}. Understanding the conditions that give rise to these pathologically coherent events, as well as the biological and physical mechanisms that regulate them, is therefore of considerable importance.

The human brain is composed of billions of neurons, which are grouped and compartmentalized. These compartments form an intricate functional connectome. In the language of graphs, each region of the brain may be modeled as a single node, and connections between regions as edges that link pairs of nodes, such that the whole brain is reduced to a complex network \cite{bullmore2009complex, li2010scale, breakspear2010generative, gerster2020fitzhugh, sawicki2024interplay} (see Fig.~\ref{fig:motivation}b). The brain connectivity is not homogeneous, and may present small-world organization \cite{li2010scale, barabasi2013network, bullmore2009complex, bassett2006small,sun2015progressive}, characterized by high connectivity and short paths between nodes. 

A variety of mathematical models exist to describe the excitable behavior of neurons and brain regions. The paradigmatic FitzHugh-Nagumo (FHN) oscillator \cite{cebrian2024six} offers a minimal framework for studying excitable systems. This two-variable model was originally introduced by FitzHugh as a modification of the Van der Pol oscillator and later demonstrated by Nagumo to be analogous to an electrical circuit \cite{fitzhugh1961impulses, nagumo1962active}. A single FHN oscillator may have excitable behavior or self-sustained oscillations. However, when coupled through a complex network topology, FHN oscillators may exhibit first or second-order phase transitions and a variety of dynamical behaviors, such as chaos and chimera states \cite{chouzouris2018chimera, scholl2016synchronization, semenova2016coherence}. In these systems, the network topology, time delays \cite{yanagita2005pair, bukh2023role}, the energy of the system \cite{li2024energy, xie2024energy}, and electromagnetic fluxes \cite{ghosh2023analysis}, among other factor, may alter the synchronization dynamics. 

Chimera states, where coherent and incoherent states coexist, are of particular relevance for biological networks \cite{lainscsek2019cortical}, as they provide a physical mechanism for partial synchronization, a feature of functioning neural circuits. Works on the FHN model have extensively studied the appearance of chimera states in non-locally coupled rings and complex networks, providing a deep understanding of the mechanisms and conditions under which chimera states emerge \cite{rontogiannis2021chimera, scholl2016synchronization, makinwa2023experimental, chouzouris2018chimera, omelchenko2015robustness, omelchenko2013nonlocal, zakharova2020chimera}. Additionally, there have been attempts to connect chimera states observed in FHN oscillators on networks to epileptic-seizures-related (ESR) synchronization in the brain \cite{andrzejak2016all,gerster2020fitzhugh}. In these systems, the dynamic is intermittent, switching back and forth between chaotic and chimera states. Biologically, these chimeras correspond to periods of abnormal synchronous activity that result in a loss of normal brain function \cite{gerster2020fitzhugh, fisher2005epileptic,borges2023intermittency}. 

In an attempt to characterize the oscillatory behavior of theoretical and real neural systems, researchers have investigated the emergence of extreme events (EEs) in specific observables, such as the oscillation amplitude or level of synchronization \cite{saha2017extreme,hernandez2021noise}. In this context, an EE occurs when the observable takes a value that is abnormally large compared to its typical behavior. Among other applications, EEs have been used to identify emergent phases in locally coupled Hindmarsh-Rose neurons \cite{sree2024extreme} and in the bursting of neuronal activity \cite{mishra2018dragon, hariharan2024noise}. Thus, extreme event analysis provides a versatile tool to characterize complex system dynamics.

Questions remain open regarding how real neural systems transit from a normal to an ESR intermittent state. Furthermore, understanding how this transition may be predicted and controlled may have a positive impact on the quality of life of people affected by related health conditions. In this work, we approach some of these questions from a theoretical perspective, using a minimal model of FHN oscillators on Watts-Strogatz (WS) complex networks \cite{barabasi2013network}. These networks capture some of the small-world properties of the brain and allow for the investigation of the influence of network randomness through the rewiring probability parameter $p$, and network size on the collective dynamics \cite{barrat2000properties} (see Fig.~\ref{fig:motivation}c and \ref{fig:motivation}d).

We show that the system of FHN oscillators on a WS network has four distinct dynamical phases: chaotic, intermittent, partially and fully synchronized. The intermittent phase exhibits transients of highly coherent, chimera states, separated in time by chaotic behavior, which were previously related to ESR events \cite{gerster2020fitzhugh}. 
We find that the network size and the level of randomness of its connectivity, as well as the coupling strength between oscillators, control the emergence, duration and time between ESR events. Furthermore, we characterize the transition into and out of the zone of existence of chimeras. By analyzing the existence of EEs in the time series of the Kuramoto global order parameter, we show that the transitions from chaos to intermittency and the transition to full synchronization are preceded by a marked increase in the probability of observing short transients of anomalously high synchronization.

The paper is organized as follows: In Secs. \ref{sec:model} and \ref{sec:Results}, we introduce the model and identify its dynamical regimes. In Sec.~\ref{sec:epilepticevents}, we show that the frequency and duration of ESR events depend on coupling strength, the number of oscillators, and network randomness. In Sec.~\ref{sec:spaceparam} we study the dynamical regimes and the stability of the synchronous solution. In Sec.~\ref{sec:ee} show that the transitions between dynamical regimes can be characterized by the likelihood of observing EEs of synchronization. Finally, in Sec.~\ref{sec:conc}, we summarize and discuss our results and their potential relevance to biological systems.

\begin{figure*}[t!]
\includegraphics[width=1.0\textwidth]{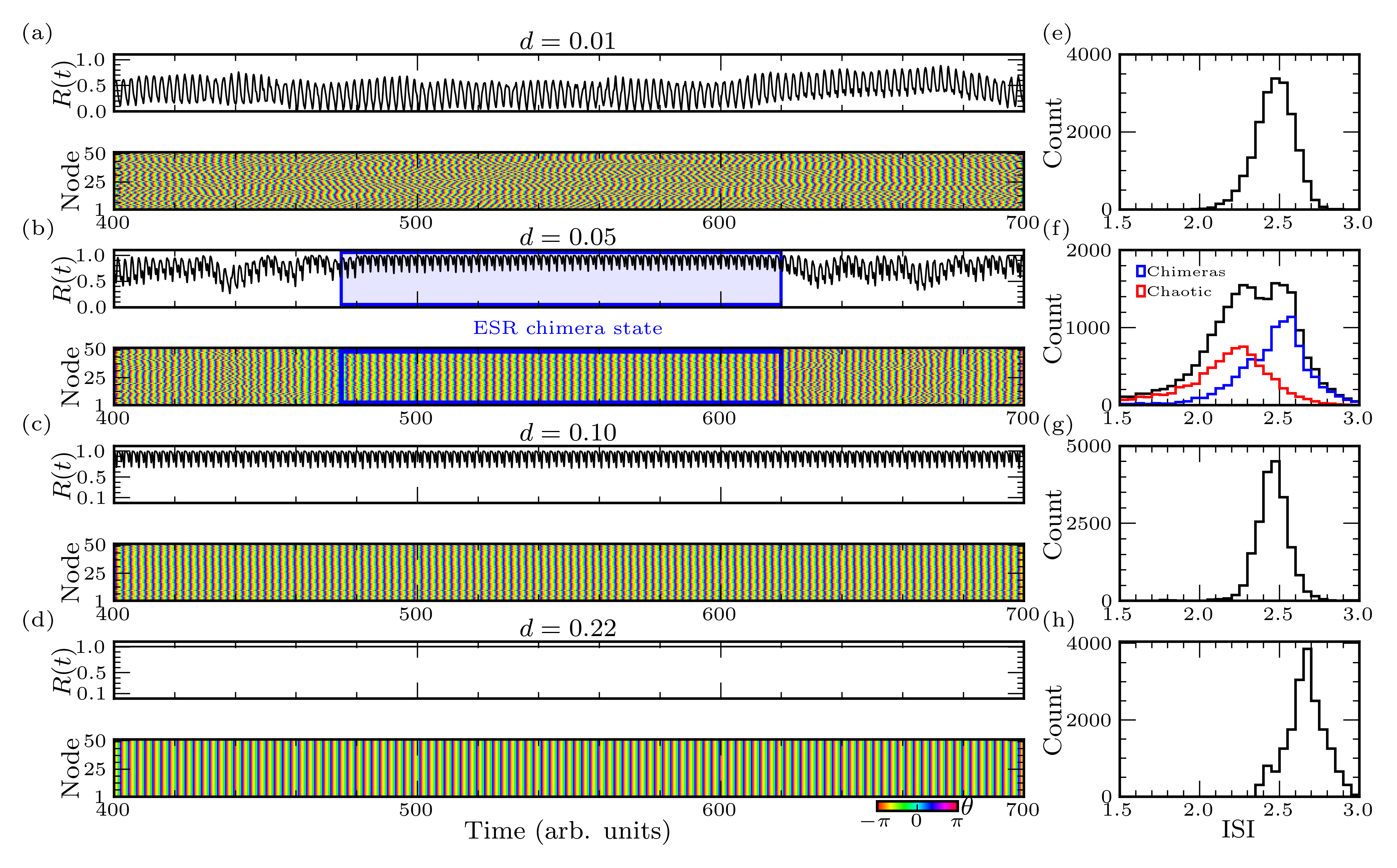}% Here is how to import EPS art
\caption{\label{fig:dynamicalpictures} \textbf{Dynamical regimes of FHN oscillators on WS networks}. (a)-(d) Show the time series of the Kuramoto order parameter $R(t)$ (top), and the temporal evolution of the phases $\theta_j(t)\in[-\pi,\pi]$ for each of the $j=1,...,N=50$ oscillators (bottom), (e)-(h) show the distributions of inter-spike intervals (ISIs) for the $N=50$ oscillators, for couplings $d=0.01$ (a,e), $d=0.05$ (b,f), $d=0.10$ (c,g) and $d=0.22$ (d,h), in WS networks with $p=1$. In (a), the time window displayed is arbitrarily chosen to illustrate the dynamic regime of interest. An ESR event is highlighted in (b). In (f), red and blue curves show the ISI distribution for the chaotic regime and chimera transients, respectively. 
}
\end{figure*}

\subsection{\label{sec:model}FHN oscillators on complex networks}

The FitzHugh-Nagumo model comprises two variables: An activation variable, $x(t)$, capturing the rapid activation dynamics, and an inhibition variable, $y(t)$, representing the slower recovery process. In this study, we consider a network of $N$ coupled FHN oscillators, where the dynamics of each oscillator is governed by \cite{scholl2009time, brandstetter2010interplay, omelchenko2013nonlocal, omelchenko2015robustness}

\begin{equation}\label{eq:coupledFHN}
\left(
  \begin{array}{c}
     \varepsilon \dot{x}_i\\
     \dot{y}_i
  \end{array}
\right ) =
\left(
  \begin{array}{c}
     x_i - \frac{x_i^3}{3}-y_i\\
     x_i + a
  \end{array}
\right ) + d \sum_{j=1}^N A^p_{ij} \boldsymbol{R}(\alpha) 
\left(
  \begin{array}{c}
     x_j-x_i\\
     y_j-y_i
  \end{array}
\right ).
\end{equation}
Here, $\dot{x}$ and $\dot{y}$ correspond to the time-derivatives of $x(t)$ and $y(t)$, respectively, and the time-scale separation $\varepsilon=0.05$ is fixed so that $x(t)$ has a fast dynamics compared to $y(t)$. The parameter $a$ controls the transition from excitable ($|a|>1$) to oscillatory ($|a|<1$) behavior. As our focus here is to study interactions of actively firing units, we set $a=0.5$. The summation term in Eq.~(\ref{eq:coupledFHN}) accounts for the inter-oscillator coupling, which is proportional to a uniform coupling strength $d$. The coupling matrix $\boldsymbol{R}(\alpha)$ is defined as
\begin{equation}
\boldsymbol{R}(\alpha) = 
\left(
  \begin{array}{cc}
     \cos\alpha & \sin\alpha \\
     -\sin\alpha & \cos\alpha 
  \end{array}
\right ),
\end{equation}
and controls the level of direct coupling (diagonal terms) and cross-coupling (off-diagonal terms). We consider $\alpha = \pi/2-0.1$, as values of $\alpha$ close to $\pi/2$ have been shown to promote the existence of chimera states \cite{omelchenko2013nonlocal}. The small-world network connectivity, for a given value of the rewiring probability $p$, is characterized by its adjacency matrix $A^p$, whose elements satisfy $A_{ij}^p = \delta_{ij}k_i - \mathcal{L}_{ij}^p$,  with $\delta_{ij}$ the Kroenecker delta, $k_i$ the degree (or number of connections) of the $i$th node, and $\mathcal{L}^p_{ij}$ the element $(i,j)$ of the Laplacian matrix of the network. 
Here we consider undirected small-world networks, so that the adjacency matrix is symmetric, and networks are constructed according to the Watts-Strogatz (WS) algorithm with rewiring probability $p$ and average degree $\langle k \rangle = 6$ \cite{watts1998collective} (see Fig.~\ref{fig:motivation}c and \ref{fig:motivation}d). The results presented here remain largely unchanged for different choices of $\langle k \rangle$. 

\subsection{Kuramoto global order parameter}

Throughout this work, in order to characterize the synchronicity of the system, we compute the Kuramoto global order parameter \cite{kuramoto1984chemical} 

\begin{equation}\label{ec:globalordpam}
    R(t) :=  \frac{1}{N}\left | \sum_{j=1}^{N} e^{i \theta_j(t) }\right |, %_{\Delta T},
\end{equation}
where $\theta_j(t)$ corresponds to a dynamical phase associated with the fast and slow variables $x_j(t)$ and $y_j(t)$ of the oscillator $j$ for $j = \left \{ 1,...,N \right \}$, defined as $\theta_j(t)~=~\arctan(y_j(t)/x_j(t))$. $R(t) \in [0,1]$, and gives relevant information about the collective dynamical regimes of the oscillators.
Values of $R(t)=1$ account for global synchronization, where $x_i = x_j$ and $y_i = y_j$ for all nodes $i$ and $j$. Additionally, values $0 < R(t) < 1$ indicates states of incoherence and partial synchronization \cite{rodrigues2016kuramoto}.

%Values of $R \approx 1$ correspond to highly synchronized states, with the limiting value $R = 1$ accounting for global synchronization, where $x_i = x_j$ and $y_i = y_j$ for all possible values of $i$ and $j$. When $R \ll 1$ the system is in an incoherent state, where oscillators do not share a common phase.

To study the phase space and the transitions between dynamical regimes as a function of the different parameters in the system, we integrate Eq.~(\ref{eq:coupledFHN}) numerically (see Appendix \ref{sec:appA} for details) and characterize the $N-d-p$ parameter space.

\section{\label{sec:Results}Results}

Networks of FHN oscillators exhibit various dynamical regimes, including chaotic (Fig.~\ref{fig:dynamicalpictures}a), intermittent (Fig.~\ref{fig:dynamicalpictures}b), partially and fully synchronized states (Fig.~\ref{fig:dynamicalpictures}c and \ref{fig:dynamicalpictures}d, respectively), which depend not only on the strength of the coupling between oscillators ($d$) but also on the system size ($N$) and the network randomness (controlled by the rewiring probability $p$). 

First, we measured the inter-spike-intervals (ISIs), defined as the time-intervals between consecutive peaks of the activation variable $x(t)$, for each oscillator. For fixed $N$ and $p$, a low (non-zero) coupling strength, $d$, results in incoherent dynamics, characterized by a unimodal distribution of ISIs (Fig.~\ref{fig:dynamicalpictures}e). As the coupling strength is increased, the system enters the intermittent regime, where the dynamics of $R(t)$ switches between chaotic and chimera states. In this regime, the distribution of ISIs shifts to a bimodal behavior. Closer inspection of the distribution revealed that the first peak is associated with the chaotic phases (red curve in Fig.~\ref{fig:dynamicalpictures}f), while the second peak corresponded to oscillations within the chimera states (blue curve in Fig.~\ref{fig:dynamicalpictures}f). As the coupling is increased further, intermittency is lost and the system reaches partially and then fully synchronized states, both of which exhibit unimodal distributions of ISIs (Figs.~\ref{fig:dynamicalpictures}g-h). In the partially synchronized regime (Figs.~\ref{fig:dynamicalpictures}c and \ref{fig:dynamicalpictures}g), chimera states persist for the whole integration time, and their ISI is consistent with that the intermittent chimeras in Fig.~\ref{fig:dynamicalpictures}f. In the fully synchronized phase all oscillators fire at their natural frequency, and the distribution of ISIs approximated a Dirac-delta function (Figs.~\ref{fig:dynamicalpictures}d and \ref{fig:dynamicalpictures}h). We note that the dispersion in Fig.~\ref{fig:dynamicalpictures}h comes from approximation errors in the dynamical phase analysis and localization fo the peaks.

\subsection{\label{sec:epilepticevents}Emergence of ESR synchronization events}

\begin{figure}[]
\includegraphics[width=0.5\textwidth]{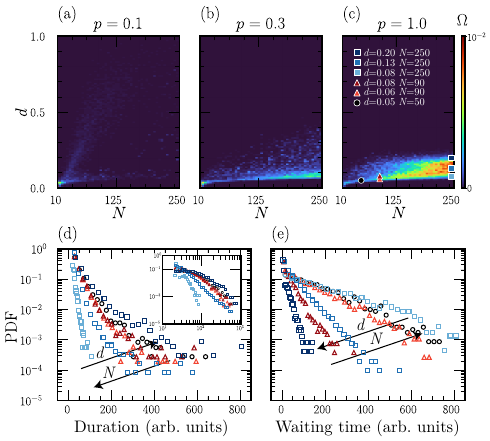}% Here is how to import EPS art
\caption{\label{fig:omega} \textbf{Emergence of ESR states}. (a)-(c) Frequency of ESR events in $(N, d)$-parameter space, for (a) $p=0.1$, (b) $p=0.3$, and (c) $p=1.0$. (d) Shows the distribution of durations of the ESR events in log-linear and (inset) log-log scale, (e) shows the distribution of waiting times between ESR events. Markers in (c) indicate the parameters analyzed in (d) and (e). Black arrows in (d) and (c) show the net effect on the distributions of increasing either $d$ or $N$, while keeping other parameters fixed.}
\end{figure}

It was previously shown \cite{gerster2020fitzhugh} that transient chimera states observed in the intermittent regime of model~(\ref{eq:coupledFHN}) resemble ESR events of synchronization in the brain (see Fig.~\ref{fig:motivation}a and \ref{fig:dynamicalpictures}b). Here, we define an ESR event as a transient chimera state whose duration is of at least $20$ simulation time units. These states are detected by analyzing the envelope of the order parameter $R(t)$ and identifying all values that lie above a threshold $R_\text{th} = 0.9$. This threshold corresponds approximately to the time-average of the order parameter $\bar{R}$ in the intermittent regime. We note that slight changes of $R_\text{th}$ do not affect our results.
With this definition, we characterize the region of existence of ESR events for a range of network sizes, $N \in [10, 250]$, rewiring probabilities, $p\in[0,1]$, and coupling strengths, $d \in [0,1]$ (Fig. \ref{fig:omega}). 

We measured numerically the frequency of ESR events ($\Omega = \text{number of ESR events}/\text{total integration time}$) as a function of the coupling strength and network size for different values of the network rewiring probability, $p$, which controls the randomness of the network (see results in Fig.~\ref{fig:omega}a-c). This analysis shows that network randomness is crucial for the emergence of ESR events, which occur only within an intermediate range of coupling strengths (see yellow/red regions in Fig.~\ref{fig:omega}b-c). While small and regular networks ($p\approx 0.1$) limit the occurrence of ESR events (Fig.~\ref{fig:omega}b), large networks with high rewiring probability have a wide range of coupling strengths for which ERS events are frequently observed (Fig.~\ref{fig:omega}c).
We then measured the distribution of durations of the ESR events and waiting times between them (see Figs.~\ref{fig:omega}d-e). The distributions of durations exhibited power-law-like decays, despite being far from the critical regime, however a more detailed analysis is required to assess their precise decay law. The waiting time between events, on the other hand, showed a clear exponential decay for all parameter values. Both properties, duration and waiting time, could be controlled by modifying the coupling strength or system size: A reduction in the coupling strength or an increase in the system size resulted in a reduction in the duration of ESR and an increase in the waiting time between them. This shows that both parameters can be used to control de properties of ESR events.

In the following, we study more generally the region of existence of chimeras (including both ESR events and persisting chimeras), and in particular, the transitions between the distinct dynamical regimes.

\subsection{\label{sec:spaceparam} Characterizing the \textit{N-d-p} phase space}

\begin{figure}[]
\includegraphics{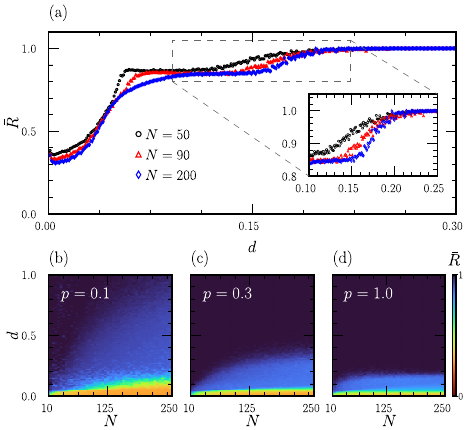}% Here is how to import EPS art
\caption{\label{fig:transitionsspaceparameter}\textbf{Phase transitions of coupled FHN oscillators}. (a) Time-averaged order parameter, $\bar{R}$, as a function of the coupling strength, $d$, for WS networks with rewiring probability $p=1$ and varied sizes, $N$. Inset shows a close-up of the transition from intermittency to global synchronization. (b)-(d) $\bar{R}$ in the $(N, d)$-parameter space, for (b) $p=0.1$, (c) $p=0.3$, and (d) $p=1.0$.}
\end{figure}

To characterize the transitions between dynamical states in the system, we compute the time-averaged global order parameter ($\bar{R}$) as a function of the coupling strength in WS networks of different sizes and $p=1$ (see Fig.~\ref{fig:transitionsspaceparameter}a). We note that regardless of the network size, $\bar{R}$ grows monotonically with $d$, exhibiting two clear transitions: One at $d \sim 0.05$ and another at $d\sim 0.15$, after which the system reaches full synchronization. Notably, for all system sized, in the range $d \sim [0.05,0.15]$, where chimeras are expected (i.e. in the intermittent and partially synchronized regime) the averaged order parameter exhibits a plateau, where $\bar{R}$ is approximately independent of $d$. Increasing the rewiring probability causes a significant decrease in the coupling threshold at which the oscillators become fully synchronized (see Fig.~\ref{fig:transitionsspaceparameter}b-c). 

To understand the changes in the transition to synchronization, in the following we perform a linear stability analysis of the synchronized solution.

\subsubsection{\label{msf}Linear stability analysis of the synchronous solution}

\begin{figure}[]
\includegraphics{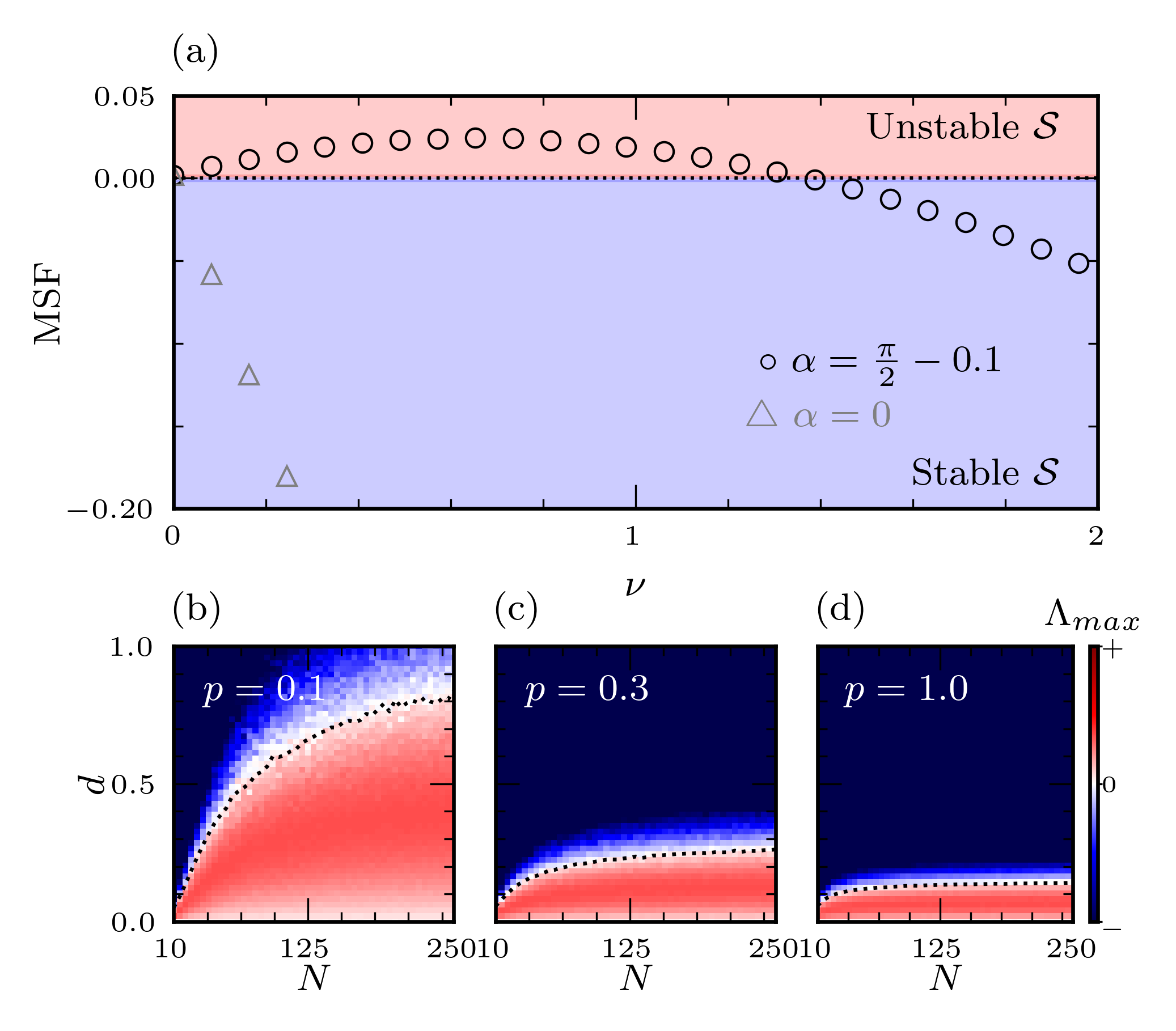}% Here is how to import EPS art
\caption{\label{fig:MSF}\textbf{MSF of FHN coupled oscillators.} (a) $\Lambda_{max}(\nu)$, for $\alpha = 0$ (gray triangles) and $\alpha = \pi/2 - 0.1$ (black circles). (b) $\Lambda_{max}$ for a WS network as a function of the system size, $N$, and coupling strength, $d$, for rewiring probabilities (b) $p=0.1$, (c) $p=0.3$, and (d) $p=1.0$. The black dashed lines in (b)-(d) show the approximated critical curve, $\Lambda_{max}(\nu) = 0$, obtained from the Wu-Chua conjecture (see Appendix~\ref{AppendixWuChua}). Shades of red and blue correspond to a positive and negative sign of the MSF, respectively.}
\end{figure}

A linear stability analysis for the synchronized solution can be achieved by evaluating the master stability function (MSF) of Eq.~(\ref{eq:coupledFHN}) \cite{pecora1998master, boccaletti2018synchronization, pikovsky2016lyapunov} (see derivation in Appendix~\ref{appendixMSF}). The MSF is defined as the maximum Lyapunov exponent $\Lambda_{max}(\nu)$ associated with the linearized dynamical system describing the time evolution of the perturbations transverse to the synchronous manifold $\mathcal{S}$. The MSF defines the stability of $\mathcal{S}$ as a function of a single parameter $\nu = \gamma_2 d$, where $d$ is the coupling strength and $\gamma_2$ is the second smallest eigenvalue of the network Laplacian matrix, known as algebraic connectivity \cite{chung1997spectral}. When $\Lambda_{max}(\nu)>0$, $\mathcal{S}$ is unstable, while $\Lambda_{max}(\nu)<0$ provides a necessary condition for the stability of $\mathcal{S}$ \cite{boccaletti2018synchronization}. In Fig.~\ref{fig:MSF}a, we show that $\Lambda_{max}(\nu)$ is positive for a range of values of $\nu$ for $\alpha = \pi/2-0.1$, however, when $\alpha = 0$ the MSF is always negative and neither chimeras nor chaotic behavior can be observed.

Consistent with the observations of the time-averaged global order parameter (Fig.~\ref{fig:transitionsspaceparameter}b-d), lowering the rewiring probability $p$ causes a shift in the critical curve $\Lambda_{max}(\nu) = 0$ towards larger values of $d$ (see Fig.~\ref{fig:MSF}b-d). We note that the critical curve was accurately predicted by the Wu-Chua conjecture (see Appendix~\ref{AppendixWuChua}), delimiting the values above which the synchronized manifold is stable (see dashed line in Fig. \ref{fig:MSF}b-d).

The MSF analysis shows that the change in the transition line to synchronization comes from changes in the topological features of the network, which enter in the MSF through the second smallest eigenvalue of the Laplacian matrix of the connectivity network. With this, we have characterized the transition from the incoherent to the coherent regime. However, the transition from chaos to intermittency is not captured by the MSF approach.

%tengo una duda sobre lo que comento a continuación: Son los estados quimeras los que nos motivan a comentar que igual se aproxima al manifold de sincronización? los eventos extremos no necesariamente son de extrema sincronización en el sentido de que se acercen a S, sino que son extremdamente sincronizados en el contexto del caos, o de la intermitencia.

As seen earlier with chimera states, the system can still approach the synchronization manifold in regions where the synchronized solution is unstable. In the following, we show that before transitioning from one dynamical regime to another, the probability that the system exhibits anomalously high levels of synchronization reaches maximum levels. We use this as a means for characterizing the region of existence of chimera states and, thus, ESR events.

\subsection{\label{sec:ee}Extreme events at the onset of phase transitions}

\begin{figure}[]
\includegraphics{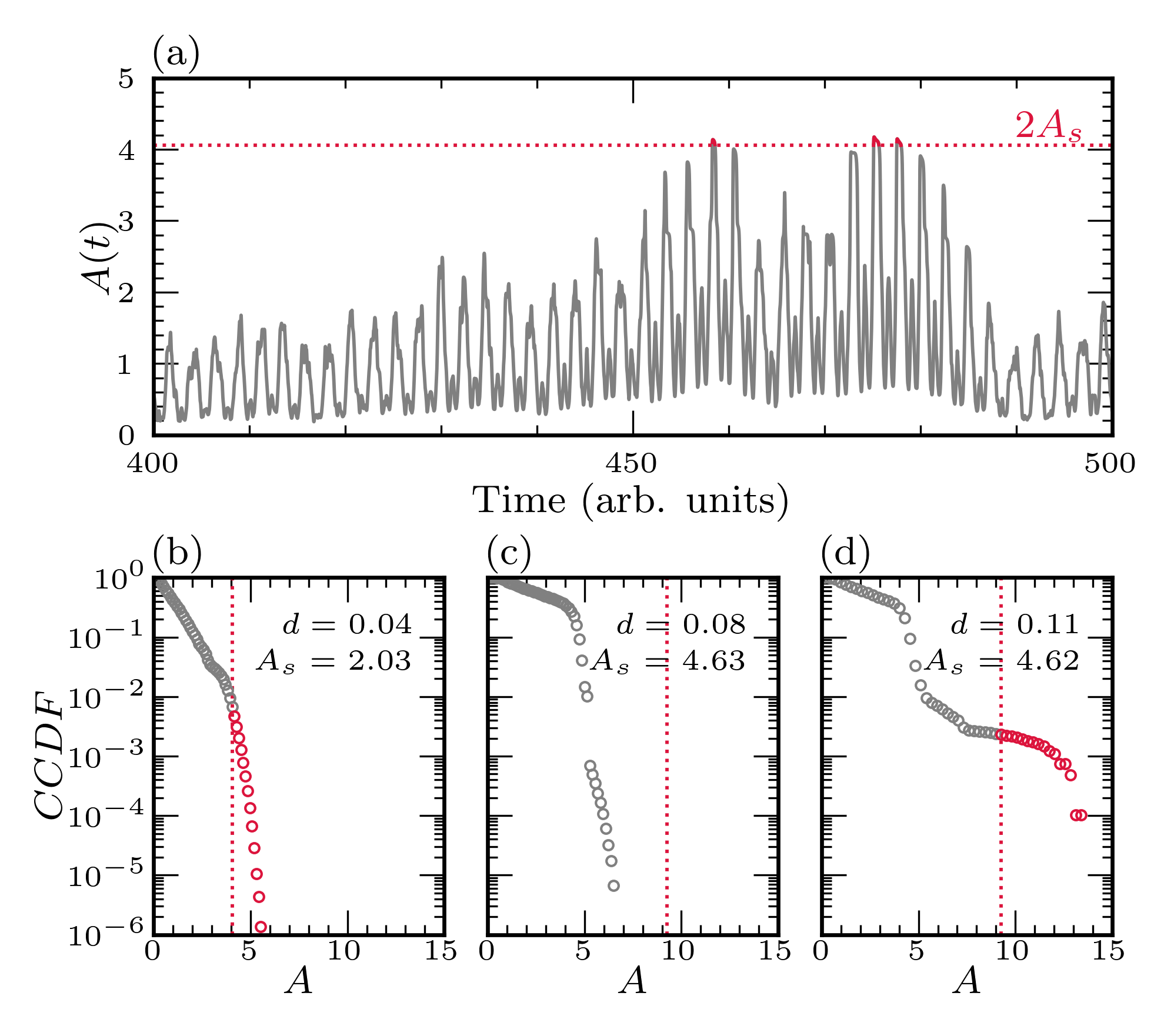}% Here is how to import EPS art
\caption{\label{fig:extremeevents} \textbf{EEs on networks of FHN oscillators}. (a) Time-series of synchronization amplitude $A(t)$, for $N=50$ oscillators, $d=0.04$ and $p=1.0$. The dotted line show the value $2 A_s$ above which the values of $A(t)$ are considered EEs (shown in red). (b)-(d) Complementary cumulative distributions of $A$, for (b) $d=0.04$, (c) $d=0.08$, and (d) $d=0.11$. The vertical dotted lines show the value $2 A_s$.}
\end{figure}

By constructing the probability distribution (PDF) of the global order parameter, we studied the likelihood that extreme events of anomalously high synchronization occur as a function of the system parameters. For this, we use the standard hydrodynamical criterion \cite{kharif2008rogue, selmi2016spatiotemporal, clerc2016extreme}: Given the PDF $P(A)$ of an observable $A$, a value $y$ of the observable is considered to be extreme if it is larger than $2A_s$, with $A_s$ defined as the mean of the third tertile of $P(A)$, i.e., events with an abnormality index $AI\equiv A/A_s > 2$ (see Fig.~\ref{fig:extremeevents}a).
To apply this criterion, we considered the observable $A(t) = -\log(1-R(t))$, which maps the domain of $R(t)$ from $[0,1]$ to $[0,+\infty)$. This ensures that the $AI$ is always contained in the domain. 
Throughout this section, we refer to $A$ as \textit{synchronization amplitude}. The proportion of EEs for a time series $A(t)$ is defined as

\begin{equation}
\begin{split}
p_{EE} = \int_{2A_s}^\infty P(A) dA.
\end{split}
\end{equation}

To illustrate how $p_{EE}$ changes with the coupling strength, we consider a system of $N=50$ oscillators in a WS network with $p=1$. As the coupling strength $d$ is increased from $d=0.04$ to $d=11$, the system transitions from chaotic ($d=0.04$), to intermittent ($d=0.08$), to partial synchronization ($d=0.11$), and the distribution $P(A)$ changes significantly (see Figs.~\ref{fig:extremeevents}b-c). In the chaotic regime and in the transition to synchronization, the system exhibits a positive proportion of extreme events (see red markers in Figs.~\ref{fig:extremeevents}b and \ref{fig:extremeevents}d). However, in the intermediate region, where ESR events are observed, the proportion of EEs is zero (see Fig.~\ref{fig:extremeevents}c). 

\begin{figure}[t!]
\includegraphics{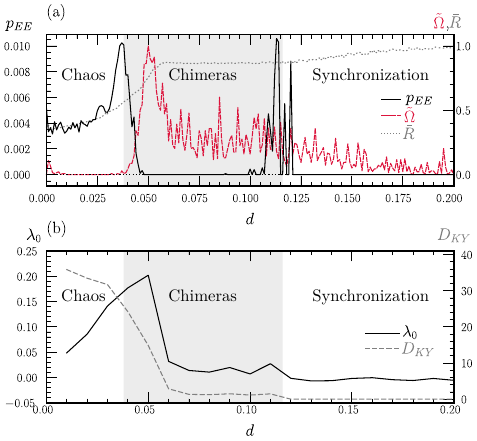}% Here is how to import EPS art
\caption{\label{fig:Chaosanalysis} \textbf{EEs and Lyapunov analysis}. (a) Proportion of EEs, frequency of ESR events (normalized by its maximum value) and order parameter (reproduced from Fig.~\ref{fig:transitionsspaceparameter}a), as a function of the coupling strength $d$. (b) Maximum Lyapunov exponent and Kaplan-Yorke dimension as a function of the coupling strength $d$. All data shown were obtained from considering a system of $N=50$ FHN oscillators on WS networks with $p=1.0$. (a-b) The shaded region indicates the existence of chimeras, the transition point from the region of chaos to chimeras is identified by the appearance of ESR events, while the transition point from chimeras to synchronization corresponds to $d=0.116$, as predicted by the MSF analysis.}
\end{figure}

A systematic analysis of the proportion of EEs revealed a non-monotonic behavior of $p_{EE}$ as the coupling strength is increased from zero up to the region of full synchronization, as shown in Fig.~\ref{fig:Chaosanalysis}a for $N=50$ and $p=1$. For low couplings ($d<0.05$) the $p_{EE}$ is always positive, showing a peak of events prior to the transition to intermittency, where no EEs are observed. As $d$ is increased further, a second peak in the $p_{EE}$ is observed (for $d \approx 0.11$), above which the system reaches full synchronization and $p_{EE}$ again falls to zero. 

To understand this behavior, we looked in more detail at the dynamic before the synchronization transition. For this, we computed the complete set of $2N$ Lyapunov exponents $\left \{ \lambda_i \right \}_{i=1}^{2N}$ for the system with $N=50$ and $p=1$, for all values of $d$ \cite{christiansen1997computing, skokos2009lyapunov, pikovsky2016lyapunov}. The maximum Lyapunov exponent $\lambda_0$ as a function of $d$ is shown in Fig.~\ref{fig:Chaosanalysis}b, which remain positive below the synchronization transition. We also computed the Kaplan-Yorke dimension, corresponding to an upper bound to the dimension of the strange attractor, which is defined as \cite{skokos2009lyapunov, pikovsky2016lyapunov}
\begin{equation}
    D_{KY} = j + \sum_{i=1}^{j} \frac{\lambda_{i}}{|\lambda_{j+1}|},
\end{equation}
where $j$ is the largest index such that $\sum_{i=1}^{j} \lambda_i \geq 0$. For a system with $N=50$ oscillators, the first region of positive $p_{EE}$ ($d<0.05$) corresponds to high-dimensional chaos (see $D_{KY}$ in Fig.~\ref{fig:Chaosanalysis}b), with the magnitude of the largest Lyapunov exponent increasing with $d$ before suffering a drastic reduction, as the peak in the $p_{EE}$ is passed (see $\lambda_0$ in Fig.~\ref{fig:Chaosanalysis}b). The peak in $p_{EE}$ is followed by a reduction of the Kaplan-Yorke dimension and marks the onset of intermittency, where ESR events are observed (see Fig.~\ref{fig:Chaosanalysis}a). Both the intermittent and the partially synchronized regime are characterized by low-dimensional chaos (with only one positive Lyapunov exponent). The second peak in the $p_{EE}$ coincides with the critical curve ($\Lambda_{max} = 0$) predicted by the MSF analysis (see Fig.~\ref{fig:MSF}d), after which the largest Lyapunov exponent and the Kaplan-Yorke dimension are further reduced. 

\begin{figure}[t!]
\includegraphics{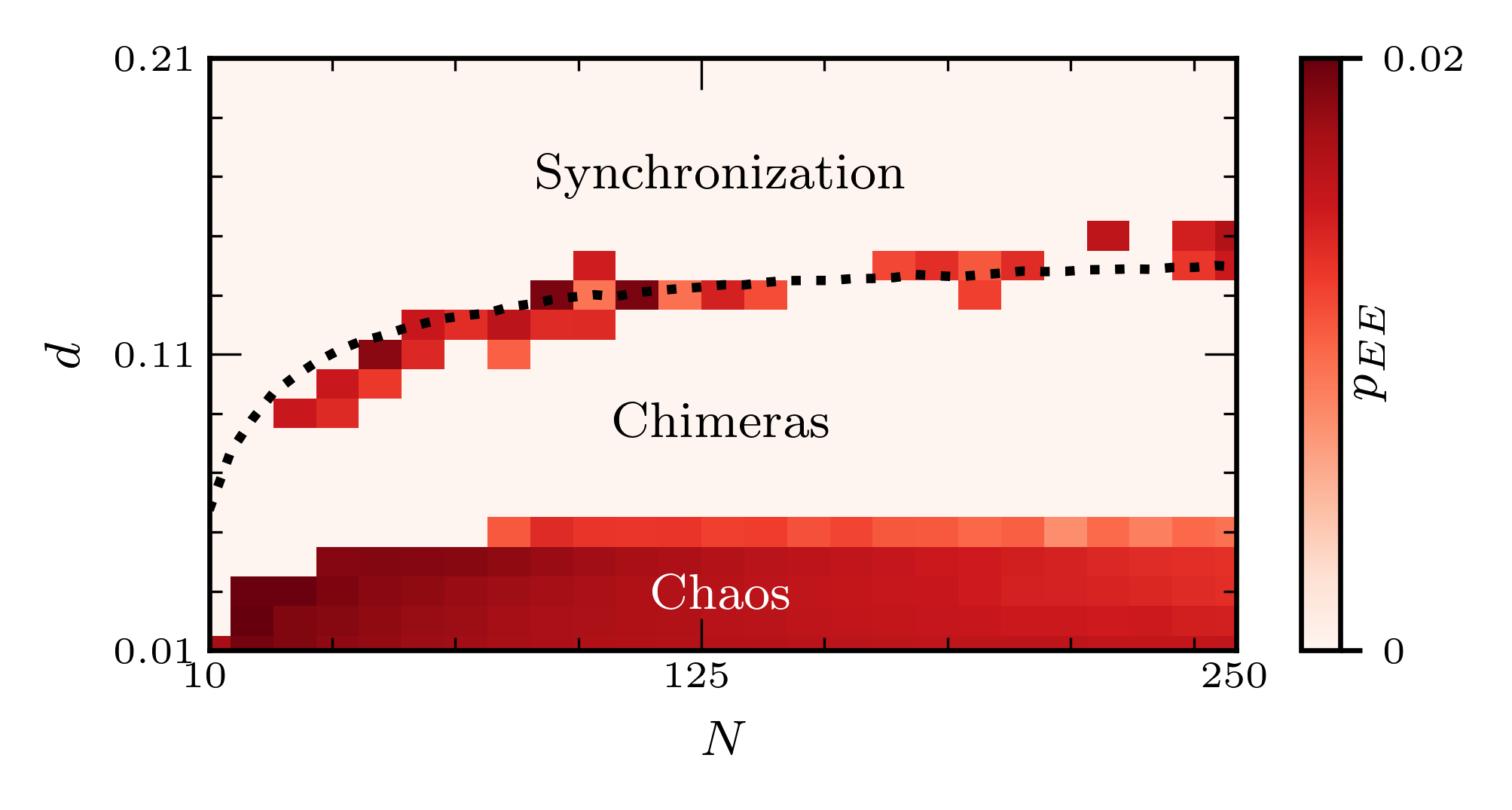}% Here is how to import EPS art
\caption{\label{fig:p_ee_space_parameter}\textbf{EEs on WS networks of FHN oscillators}. Proportion of EEs ($p_{EE}$) as a function of the system size $N$ and coupling strength $d$, for $p=1.0$. Pale color indicates regions where $p_{EE} = 0$. The dashed line shows the critical curve where the MSF changes sign, obtained through the Wu-Chua conjecture.}
\end{figure}

%Comparing the time-averaged global order parameter, frequency of ESR events, and the proportion of EEs (Figs.~\ref{fig:Chaosanalysis}a-b), we observe that changes in $D_{KY}$, as well as the disappearance and appearance of EEs, are directly correlated with the critical values of $d$ where the transitions occur. 
We then computed the $p_{EE}$ in the complete $N-d$ parameter space (see Fig.~\ref{fig:p_ee_space_parameter}). Here, we notice that in the chaotic regime $p_{EE}$ is always positive. As noted before, at the onset of intermittency, the proportion of EEs exhibits a peak, decaying to zero once the intermittent region is reached. By further increasing the coupling strength, a second peak is observed in the proportion of EEs, accurately matching the transition curve obtained from the MSF analysis for the stabilization of the synchronization manifold. We note that our analysis based on the proportion of EEs does not distinguish between the intermittent regime (i.e. coexistence of chaos and chimeras) and that of partial synchronization (in which chimeras persist for the whole duration of the simulation). %This is possibly due to the fact that the transition between these regimes occurs gradually as $d$ is increased, which causes an increase in the duration of chimeras and the reduction in the waiting time between them (see Figs.~\ref{fig:omega}d and \ref{fig:omega}e). 
 
\section{\label{sec:conc}Conclusions and discussion}

In this study, we explored the dynamics of coupled FitzHugh-Nagumo oscillators on Watts-Strogatz small-world networks. Our results show that the network randomness and coupling strength significantly influence the occurrence of ESR states, which resemble the transient synchronization observed in biological neural networks during absence epileptic seizures. 
%aqui voy
Interestingly, ESR events occur only within an intermediate range of coupling strengths and become more frequent and of shorter duration as the network becomes larger, provided that the rewiring probability is large enough (as shown in Fig.~\ref{fig:omega}b-c). This opens questions about the compartmentalization in animal brains and the mechanisms of wiring in healthy and altered conditions. In humans, for instance, anatomical inspection of the brain enables the identification of approximately 90 distinct regions \cite{tzourio2002automated, vskoch2022human}. According to our theoretical observations, this brain organization promotes the coexistence of coherent and incoherent phases, however, it remains unclear which are the mechanisms that lead to this specific organization.  

Evaluating the master stability function allowed us to find the transition curve to synchronization in the $N-d$ parameter space for different values of the rewiring probability $p$. This curve separated the phase space into a synchronized and a chaotic phase, as demonstrated by the Lyapunov exponent analysis. Within the chaotic phase, we distinguished two regions, which became apparent when analyzing the proportion of EEs in the (rescaled) order parameter: A fully chaotic region and a region of coexistence between an incoherent and a coherent phase, where both chimera states and ESR events were observed. We have shown that the transitions from chaos to intermittency, and subsequently to synchronization, are preceded by a peak in the frequency of extreme synchronization events, a behavior that correlates with significant changes in the dimensionality of the system's strange attractor.
%lo ultimo que comenté creo que puede confundirse con los epileptic-like events. En el párrafo de abajo, quizás sería bueno incluir algo de criticalidad (?

In general, computing the proportion of EEs $p_{EE}$ requires only a sufficiently long time series in order to construct the PDF of a relevant observable, which may provide useful information of the distinct dynamical phases of the system. This type of characterization of the phase space could open the door for the potential analysis of real EEG data, where time series of neural dynamics are more readily available. 

In conclusion, this work contributes to a deeper understanding of the interplay between network structure, coupling strength, and neural dynamics. The findings may have broader implications for understanding neural processes, offering potential applications in modeling the conditions that lead to pathological states such as ESR intermittency and synchronous states. Future work could extend this analysis to dynamical network topologies and explore the effects of additional biological factors such as noise and time delays, to better model realistic neural systems. 

\begin{acknowledgments}
I.B. acknowledges the financial support from FONDECYT grant 11230941, and Universidad de Chile-VID grant UI-015/22. The authors wish to thank Magdalena Sanhueza, Claudio Falcón, Karin Alfaro-Bittner and Marcel Clerc for useful discussions.
\end{acknowledgments}

\appendix

\section{\label{sec:appA}Numerical simulations}

The dynamics of coupled FHN oscillators are solved by integrating Eq.~(\ref{eq:coupledFHN}) with a fourth-order Runge-Kutta method, considering random initial conditions following a uniform distribution $x \sim U_x(-a, a)$, $y \sim U_y(-a + a^3/3, a + a^3/3)$ and a suitable time step that ensures convergence. Specifically, simulations in (Fig. \ref{fig:dynamicalpictures}) consider the time evolution of $10^4$ steps, and the ISI PDF is calculated after a transient of $10^2$ time steps. Simulations in (Fig. \ref{fig:omega}a-e, \ref{fig:transitionsspaceparameter}a-d and \ref{fig:Chaosanalysis}a) consider the time evolution of $10^5$ time steps. The computation of $\Omega$ for each point in the $N-d-p$ space in (Fig. \ref{fig:omega}a-c) is calculated by averaging the analysis of 50 realizations of $R(t)$, excluding the final $10^3$ time steps, in order to consider only ESR events. Additionally, distributions in (Fig. \ref{fig:omega}d-e) are calculated based on the analysis of 1500 realizations of $R(t)$, for each system size and coupling strength indicated. The time-averaged global order parameter $\bar{R}$ in (Fig.\ref{fig:transitionsspaceparameter}a-d, \ref{fig:Chaosanalysis}a) is calculated as a long-time average of $5 \times 10^4$ time steps after a sufficiently large transient of $5 \times 10^4$ time steps, averaged over 50 realizations. The proportion of EEs $p_{EE}$ in (Fig.\ref{fig:Chaosanalysis}a) is calculated through the analysis of 50 realizations of $A(R(t))$. Simulations in (Fig. \ref{fig:extremeevents}a-d and \ref{fig:p_ee_space_parameter}) consider the time evolution of $10^6$ time steps, and both the distributions and $p_{EE}$ are calculated considering 50 realizations of $A(R(t))$. Finally, the dynamical systems involved in the algorithms for computing the Lyapunov exponents \cite{christiansen1997computing, skokos2009lyapunov, pikovsky2016lyapunov} were solved with a fourth-order Runge-Kutta method for $10^3$ time units with an adaptive time step and suitable parameters that ensure convergence.

\section{\label{appendixMSF}Master Stability Function}

To perform the linear stability analysis for understanding the transition to synchronization, we study the MSF of the system of Eq.~(\ref{eq:coupledFHN}) as in \cite{boccaletti2018synchronization,rakshit2018emergence}. Let $(x_S, y_S)^T \in \mathcal{S}$, whose orbits obey:

\begin{equation}
\begin{split}
\varepsilon \dot{x}_S(t) &= f(x_S, y_S) = x_S - \frac{x_S^3}{3} - y_S \\
\dot{y}_S(t) &= g(x_S, y_S) = x_S + a.
\end{split}
\end{equation}

Let $(\delta x_i, \delta y_i)^T = (x_i(t) - x_S, y_i(t) - y_S)^T$ be small perturbations away from $\mathcal{S}$. Consider now the relation $A_{ij}^p = \delta_{ij} k_i - \mathcal{L}_{ij}^p$, where $\delta_{ij}$ is the Kronecker delta, $k_i$ is the degree of the $i$ node, $p$ is the rewiring probability of the Watts-Strogatz model and $A_{ij}^p, \mathcal{L}_{ij}^p$ are the elements of the adjacency and Laplacian matrix of the network $A^p$ and $\mathcal{L}^p$, respectively. Linearizing around $(x_S, y_S)^T$, and using the property $\sum_j \mathcal{L}_{ij}^p = 0$, the perturbations evolves as

\begin{equation}\label{eq:AppBMSF1}
\begin{split}
 \dot{\delta x_i} &= \frac{(1 - x_S^2)}{\varepsilon} \delta x_i + \frac{\delta y_i}{\varepsilon} - d\sum_{j=1}^{N}\mathcal{L}_{ij}^p\left[ \frac{\cos\alpha}{\varepsilon}\delta x_j\ + \frac{\sin\alpha}{\varepsilon}\delta y_j\right]\\
\dot{\delta y_i} &= \delta x_i - d\sum_{j=1}^{N}\mathcal{L}_{ij}^p\left[ -\sin\alpha\delta x_j + \cos\alpha\delta y_j \right].
\end{split} 
\end{equation}

\noindent Now let $V^p$ be the matrix of eigenvectors of $\mathcal{L}^p$ and $\bm{{\xi}}_i = (\xi_{i}^{x}, \xi_{i}^{y})^T$ the projections of the perturbations on the eigenvector basis

\begin{equation}\label{}
    \bm{{\xi}}_i = (\xi_{i}^{x}, \xi_{i}^{y})^T = \Bigl( \sum_{j=1}^{N} V_{ij}^p \delta x_j,  \sum_{j=1}^{N} V_{ij}^p \delta y_j\Bigl)^T,
\end{equation}

\noindent which corresponds to the decoupled eigenmodes identified by the eigenvectors of $\mathcal{L}^p$ with associated eigenvalues $\gamma_1^p < \gamma_2^p < ... < \gamma_i^p < ... < \gamma_N^p$, thus, the relation $\mathcal{L}_{kj}^p V_{ij}^p = \delta_{kj} V_{ij}^p \gamma_i^p$ holds. Using this, Eq.~(\ref{eq:AppBMSF1}) becomes

\begin{equation}\label{}
\begin{split}
 \dot{\xi_{i}^{x}} &= \frac{(1 - x_S^2)}{\varepsilon}\xi_{i}^{x} -\frac{\xi_{i}^{y}}{\varepsilon} - \gamma_i^p d  \left[\frac{\cos\alpha}{\varepsilon}\xi_{i}^{x} + \frac{\sin\alpha}{\varepsilon}\xi_{i}^{y}\right]\\
 \dot{\xi_{i}^{y}} &= \xi_{i}^{x} - \gamma_i^p d  \left[-\sin\alpha\xi_{i}^{x} + \cos\alpha\xi_{i}^{y} \right].
\end{split}
\end{equation}

\noindent The equations above can be written as

\begin{equation}\label{}
\bm{\dot{\xi}}_i = \mathbf{K}(\nu_i)\bm{\xi}_i,
\end{equation}

\noindent where $\nu_i = \gamma_i^p d$, and

\begin{equation}\label{}
\mathbf{K}(\nu_i) = \begin{pmatrix}
(1 - x_{\mathcal{S}}^2 - \nu_i \cos \alpha)/\varepsilon & -1 - \nu_i \sin \alpha\\ 
(1 + \nu_i \sin \alpha)/\varepsilon & - \nu_i \cos \alpha
\end{pmatrix},
\end{equation}

\noindent letting $\nu_i$ be an arbitrary value $\nu$, we arrive to

\begin{equation}\label{eq:MSF}
\bm{\dot{\xi}} = \mathbf{K}(\nu)\bm{\xi}.
\end{equation}

\noindent We compute the MSF of the system, which corresponds to the maximum Lyapunov exponent $\Lambda_{max}$ of Eq.~(\ref{eq:MSF}) via the standard procedure developed in \cite{skokos2009lyapunov, pikovsky2016lyapunov} for suitable parameters that ensure the convergence of the algorithm. 

\section{\label{AppendixWuChua}The Wu-Chua conjecture}

To quantify and understand in a more detailed manner the stability regions of the synchronization manifold in the $N-d$ parameter space, we test the Wu-Chua conjecture \cite{wu1996conjecture}, which gives a relation between the second smallest eigenvalue of the Laplacian matrix of a graph and the critical coupling strength needed for synchronization of two different systems. Let $(s.i)$ and $(s.ii)$ be two linearly coupled systems with $N_1$ and $N_2$ units with respective Laplacian matrices $\mathcal{L}_1$ and $\mathcal{L}_2$ whose smallest eigenvalues are $\gamma_2(N_1)$ and $\gamma_2(N_2)$, then the following relation holds:

\begin{equation}\label{eq:wuchua}
    d^c_{N_1} = \frac{d_{N_2}^c \gamma_2(N_2)}{\gamma_2(N_1)},
\end{equation}

\noindent where $d^c_{N_1}$ and $d_{N_2}^c$ correspond to the critical coupling strength for synchronization for $(s.i)$ and $(s.ii)$, respectively. We consider the system $(s.i)$ and $(s.ii)$ both to be FitzHugh-Nagumo coupled oscillators whose dynamics obeys Eq.~(\ref{eq:coupledFHN}). In this case, $(s.i)$ corresponds to the system of $N$ units interacting over a Watts-Strogatz network of the model described in Sec. \ref{sec:Intro} with Laplacian matrix $\mathcal{L}_{ij}^p$, whose second smallest eigenvalue $\gamma_2(N;p, \langle k \rangle)$ is computed numerically, and $(s.ii)$ corresponds to a system of $N_2=2$ coupled units whose Laplacian matrix is $\mathcal{L}_2=\begin{bmatrix}
    1 & -1\\ -1 & 1
    \end{bmatrix}$, with second smallest eigenvalue $\gamma_2(N_2) = 2$. The synchronization threshold for system $(s.ii)$ is obtained via the MSF procedure described in Appendix \ref{appendixMSF}, and we find it to be $d^c_{N_2} \approx 0.105 \approx 1/10$. The relation (\ref{eq:wuchua}) becomes

\begin{equation}\label{eq:wuchua2}
    d^c_{WC}(N;p,\langle k \rangle) = \frac{1}{5\gamma_2(N;p,\langle k \rangle)},
\end{equation}
which correspond to the curves shown in Fig. \ref{fig:transitionsspaceparameter}b-d and \ref{fig:MSF}b-d, where we have considered $\langle k \rangle=6$ according to the model studied.

\bibliography{library}

\end{document}